\documentstyle[12pt,a4]{article}
\begin{document}
\thispagestyle{empty}
\begin{flushright}
CERN-TH/95-343\\
BI-TP 95/42
\end{flushright}
\vskip3cm

\vskip0.8cm
\centerline{\large{\bf{CAN GLUONS TRACE BARYON NUMBER ?}}}
\vskip2cm
\centerline{D. KHARZEEV}
\vskip0.3cm
\centerline{{\it Theory Division, CERN, CH-1211 Geneva, Switzerland }}
\centerline{{\it and}}
\centerline{{\it Fakult\"at f\"ur Physik, Universit\"at Bielefeld, 
D-33501 Bielefeld,
 Germany}}
\vskip1.5cm
\begin{abstract}
QCD as a gauge non-Abelian theory 
imposes severe constraints on the structure of the baryon wave function. 
We point out that, contrary to a widely accepted belief, 
the traces of baryon number in a high-energy process can reside in a 
non-perturbative configuration of gluon fields, rather than 
in the valence quarks.
We argue that this   
conjecture  
can be tested experimentally, since it can lead to substantial 
baryon asymmetry 
in the central rapidity region of ultra-relativistic 
nucleus-nucleus collisions.   
\end{abstract}
\vskip3cm
\begin{flushleft}
CERN-TH/95-343\\
BI-TP 95/42
\end{flushleft}

\newpage

In QCD, quarks carry colour, flavour, electric charge and isospin. It seems 
only natural to assume that they also trace baryon number. However, 
this latter 
assumption is not dictated by the structure of QCD, and therefore does not 
need to be true. Indeed, the assignment of the baryon number $B=1/3$ to quarks 
is based merely on the naive quark model classification. But any physical 
hadron state in QCD should be represented by a state vector which is 
 gauge-invariant -- the constraint which 
is ignored in most of the naive quark model formulations. 
This constraint turns 
out to be very severe; in fact, there is only one way to 
construct a gauge-invariant state vector of a baryon from 
quarks and gluons \cite{Ven} (note however that there is a large amount 
of freedom in choosing the paths connecting $x$ to $x_i$):
$$
B=\epsilon^{ijk}\left[P exp\left(ig\int_{x_1}^x A_{\mu}dx^{\mu}\right) 
q(x_1)\right]_i \left[P exp\left(ig\int_{x_2}^x A_{\mu}dx^{\mu}\right) 
q(x_2)\right]_j 
$$
\begin{equation}
\hskip6.5cm \times \left[P exp\left(ig\int_{x_3}^x A_{\mu}dx^{\mu}\right) 
q(x_3)\right]_k. \label{wf}
\end{equation}
The ``string operators" in (\ref{wf}) acting on the quark field 
$q(x_n)$ make it 
transform as a quark field at point $x$ instead of at $x_n$. The $\epsilon$ 
tensor then constructs a local colour singlet and gauge invariant 
state out of three quark fields (see Fig.1a). 
The $B$ in eq. (\ref{wf}) is a set of gauge invariant 
operators representing a baryon in QCD. With properly optimised parameters 
it is used extensively in the first principle computations with lattice 
Monte Carlo attempting to determine the nucleon mass. The purpose of 
this work is to study its phenomenological impact on baryon number production 
in the central region of nucleus-nucleus collisions. 
\vskip0.3cm

It is evident from the 
structure of (\ref{wf}) that the trace of baryon number should be 
associated not 
with the valence quarks, but with a non-perturbative configuration 
of gluon fields located at the point $x$ - the  
``string junction" \cite{Ven}. This can be nicely illustrated 
in the string picture: let us pull all of the quarks  
away from the string junction, which we keep fixed at point $x$. 
This will lead to $\bar{q}q$ pair production 
and string break-up, but the baryon will always restore itself around 
the string junction. The quark composition 
of this resulting baryon will in general differ from the composition of the 
initial baryon. It is important to note that the assignment of the trace of 
baryon 
number to gluons is not only a feature of a particular kind of the string 
model, but is a consequence of the local gauge invariance principle 
applied to baryons. 
\vskip0.3cm

Surprisingly, at first glance this non-trivial structure of baryons in QCD 
does not seem 
 to reveal itself in hadronic interactions in any 
appreciable way. This is because to probe the internal structure of 
baryons, we need hard interactions, which usually act on single quarks. 
The baryon number therefore can always be associated with the remaining 
diquark, without specifying its internal structure. To really understand 
 what traces the baryon number, one needs therefore 
to study processes in which the flow of baryon number can be separated 
from the flow of valence quarks.
\vskip0.3cm

In this Letter we suggest that studies of baryon production in the 
central rapidity region of ultra-relativistic $pp$ and especially $AA$ 
collisions provide a crucial possibility to test the baryon structure. 
Our findings may prove important for understanding the 
properties of dense QCD matter produced in $AA$ collisions at high 
energies. In what follows, we shall first formulate our ideas qualitatively, 
and then give some quantitative estimates based on the topological expansion 
of QCD \cite{Ven, Ven1} and Regge phenomenology.
\vskip0.3cm

Let us consider first an ultra-relativistic $pp$ collision in its 
centre-of-mass frame, which coincides with the lab frame 
in collider experiments. 
At sufficiently high 
energies, the valence quark distributions will be Lorentz-contracted to thin 
pancakes with the thickness of
\begin{equation}
z_V \simeq {1 \over x_V P}, \label{pan}
\end{equation}
where $P$ is the c.m. momentum in the collision, and $x_V \sim 1/3$ is a 
typical fraction of the proton's momentum carried by a valence quark. 
The typical time needed for the interaction of valence quarks from 
different protons with each other during the collision is given by the 
characteristic interquark distance in the impact parameter plane, 
$t_{int} = const \sim O(1\ fm)$. However, the time available for this 
interaction in the collision is only $t_{coll} \sim z_V \sim (x_V P)^{-1}$. 
It is therefore clear that at sufficiently high energies, when $t_{coll}<<
t_{int}$, the valence quarks of the colliding protons do not have time 
to interact during the collision and go through each other, populating the 
fragmentation regions. In the conventional picture, the baryon number 
follows the valence quarks. 
\vskip0.3cm

At first glance, the argument looks correct, and is well supported 
experimentally - the leading effect for baryons in high energy 
$pp$ collisions is well established.
However, the structure of the gauge-invariant baryon wave function (1) 
suggests that this scenario may not be entirely consistent. 
As we have stressed above, in QCD the trace of the baryon number 
has to be associated with the non-perturbative configuration of 
the gauge field. This ``string junction" contains an infinite number of 
gluons which, therefore, by virtue of momentum conservation, 
should carry on the average an infinitely 
small fraction $x_S<<x_V$ of the proton's momentum. We therefore expect that 
the ``string junction" configuration may not be Lorentz-contracted to 
a thin pancake even at asymptotically high energies (see Fig.2), since
\begin{equation}
z_S \simeq {1 \over x_S P} >> z_V. \label{zs}
\end{equation}
In this case the string junction will always have enough time to interact, 
and we may expect to find stopped baryons in the central rapidity region 
even in a high-energy collision (of course there will also be 
baryon-antibaryon pair production in addition). 
This argument leads to a peculiar picture of a high-energy $pp$ collision: 
in some events, one or both of the string junctions are stopped in the 
central rapidity region, whereas the valence quarks are stripped-off and 
produce three-jet events   
in the fragmentation regions. Immediately after collision, the central region 
is then filled by a gluon sea containing one or two twists, 
which will later on be dressed up by sea quarks and will form baryon(s). 
Note that the quark composition of the produced baryons will in general differ 
from the composition of colliding protons. 
\vskip0.3cm

Why then is the leading baryon effect a gross feature of high-energy 
$pp$ collisions? The reason may be the following. 
The string junction, connected to all three of the valence quarks, 
is confined inside the baryon, 
whereas $pp$ collisions become on the average more and more peripheral 
at high energies. Therefore, in a typical high-energy collision, 
the string junctions of the colliding baryons pass far away from each other 
in the impact parameter plane and do not interact.  One can however select 
only central events, triggering on high multiplicity of the produced 
hadrons. In this case, we expect that the string junctions will interact 
and may be stopped in the central rapidity region. 
This should lead to the baryon asymmetry in the central rapidity region:
even at very high energies, there should be more baryons than antibaryons 
there. 
\vskip0.3cm

Fortunately, the data needed to test this 
conjecture already exist: the experimental study of baryon and antibaryon 
production 
with trigger on associated hadron multiplicity has been already 
performed at ISR, 
at the highest energy ever available in $pp$ collisions \cite{Bel77}. 
This study has revealed that in the central rapidity region, 
the multiplicities associated with a proton are higher than with an antiproton 
by $\simeq 10\%$. It was also found that the number of baryons 
in the central rapidity region substantially exceeds the 
number of antibaryons \cite{Alp}. These two observations combined 
 indicate the existence of an appreciable baryon stopping 
in central $pp$ collisions even at very high energies \cite{Bel77}. 
\vskip0.3cm

Where else do we encounter central baryon-baryon collisions? In a 
high energy nucleus-nucleus collision, the baryons in each of the colliding 
nuclei are densely packed in the impact parameter plane, with an average 
inter-baryon distance 
\begin{equation}
r \simeq \rho^{-1/2} A^{-1/6}, \label{r}
\end{equation}
where $\rho$ is the nuclear density, and $A$ is the atomic number. The   
impact parameter $b$ in an individual baryon-baryon interaction in the 
nucleus-nucleus collision is therefore effectively cut off by the 
packing parameter: 
$b \leq r$. In the case of a lead nucleus, for example, $r$ appears to be 
very small: $r\simeq 0.4\ fm$, and a central lead-lead collision should 
therefore be accompanied by a large number of interactions among the 
string junctions. 
This may lead to substantial baryon stopping even at 
RHIC and LHC energies.
\vskip0.3cm

We shall now proceed to more quantitative considerations. In  
the topological expansion scheme \cite{Ven}, 
the separation of the baryon number flow from the flow of valence quarks 
in baryon-(anti)baryon interaction can be represented through a $t$-channel 
exchange 
of the quarkless junction-antijunction state with the wave function 
given by 
$$
M_0^J = \epsilon_{ijk} \epsilon^{i'j'k'}  
\left[P exp\left(ig\int_{x_1}^{x_2} A_{\mu}dx^{\mu}\right)\right]_{i'}^i
\left[P exp\left(ig\int_{x_1}^{x_2} A_{\mu}dx^{\mu}\right)\right]_{j'}^j
$$
\begin{equation}
\hskip6.7cm \times \left[P exp\left(ig\int_{x_1}^{x_2} A_{\mu}dx^{\mu}\right)
\right]_{k'}^k. 
\label{M}
\end{equation}
The structure of the wave function (\ref{M}) is illustrated in Fig.1b - 
it is a quarkless closed string configuration composed from a junction 
and an antijunction. In the topological expansion scheme, the states (\ref{M}) 
lie on a Regge trajectory; its intercept can be related to the 
baryon and reggeon intercepts \cite{Ven}: 
\begin{equation}
\alpha_0^J(0) \simeq 2 \alpha_B(0) - 1 + 3 (1 - \alpha_R(0)) \simeq 
{1 \over 2}, 
\label{int}
\end{equation}
where the baryon intercept $\alpha_B(0)$ has been set equal to $0$, and the 
reggeon intercept $\alpha_R(0)$ equal to  $1/2$.
\vskip0.3cm

The $M_0^J$ exchange should dominate the proton-antiproton 
annihilation at high energies \cite{Ven}. Indeed, annihilation requires 
the baryon 
number transfer in the $t-$channel. 
Conventionally, this corresponds to 
the baryon exchange with the intercept $\alpha_B(0)\simeq 0$.
 Since in Regge theory the energy dependence of the cross section is given by 
$s^{\alpha(0)-1}$, and $\alpha_0^J(0) > \alpha_B(0)$, the $M_0^J$ exchange  
should give the dominant contribution at high energies (see Fig.3), 
leading to the 
following energy dependence of the annihilation cross section:
\begin{equation}
\sigma_{\bar{p}p}^{ann} \sim \left({s \over s_0}\right)^{\alpha_0^J(0)-1} 
\simeq \left({s \over s_0}\right)^{-1/2} \label{ann}
\end{equation}
instead of the $s^{-1}$ dependence implied by conventional baryon exchange 
($s_0\simeq 1\ GeV^2$ is the usual parameter of Regge theory). 
Up to the highest energies where the annihilation can still be experimentally 
distinguished from other inelastic processes, the energy 
dependence (\ref{ann}) 
is confirmed by the data. Moreover, the entire difference between 
the total $pp$ and $\bar{p}p$ cross sections can be attributed to annihilation 
(see \cite{Ben} for a recent review). 
\vskip0.3cm

Let us now turn to the consideration of baryon stopping in $pp$ collisions. 
The relevant diagrams are shown in Fig.4; we consider 
the simultaneous stopping 
of the two string junctions in the central rapidity region, accompanied 
by three-jet events in the fragmentation regions (Fig.4a), and the stopping 
of the junction of one proton in the soft parton field of the other, 
accompanied by one three-jet event  
(Figs.4b,c).  
To calculate the cross sections, it is convenient 
 to evaluate the discontinuity of the corresponding three-particle 
elastic scattering process 
\cite{Mue,Kan} (see Fig.5).  Introducing the four-momentum $p_B$ 
of the produced 
baryon (or the total momentum of the pair of baryons in the diagram of 
Fig.5a)  and the 
four momenta of the colliding protons $p_1$, $p_2$ we have the 
following expressions for the invariant energy in the proton-baryon systems:
$$
s_1 \simeq \sqrt{s}\ m_t\ e^{-y^*}, 
$$  
\begin{equation}
s_2 \simeq \sqrt{s}\ m_t\ e^{y^*}, \label{kin} 
\end{equation}  
where $s=(p_1+p_2)^2$ is the c.m.s. energy squared of the $pp$ collision, 
$y^*$ is the c.m.s. rapidity of the produced baryon(s), and $m_t$ is its 
transverse mass $m_t^2=m_B^2+p_B^2$. The product of the invariants (\ref{kin}) 
satisfies the relation
\begin{equation}
s_1 s_2 \simeq m_t^2 s. \label{prod}
\end{equation}
Let us denote the coupling of the $M_0^J$ reggeon and pomeron to the proton by 
$G_p^M$ and $G_p^P$ respectively, and 
introduce scalar functions $f_B^{MM}(m_t^2)$ and $f_B^{MP}(m_t^2)$ 
describing the ``two baryons - $M_0^J\ - M_0^J$" and 
``one baryon - $M_0^J$\ - Pomeron" vertices. The standard calculation 
\cite{Mue,Kan} then allows us to calculate the cross sections; for    
the diagram of Fig.5a we get
\begin{equation}
E_B {d^3 \sigma^{(2)} \over d^3 p_B} = 8 \pi [G_p^M(0)]^2\ f_B^{MM}(m_t^2)\ 
\left({\sqrt{s}\ m_t \over s_0}\right)^{2\alpha_0^J(0) -2}. \label{cross2}
\end{equation} 
Analogous calculation for the sum of diagrams in Figs.5b,c gives
$$
E_B {d^3 \sigma^{(1)} \over d^3 p_B} = 8 \pi G_p^M(0) G_p^P(0)\ 
f_B^{MP}(m_t^2)\ 
\left({\sqrt{s}\ m_t \over s_0}\right)^{\alpha_0^J(0) + \alpha_P(0) -2} 
$$
\begin{equation}
\hskip2cm \times \left(exp[y^*(\alpha_P(0) - \alpha_0^J(0))] + 
exp[-y^*(\alpha_P(0) - \alpha_0^J(0))] 
\right). \label{cross}
\end{equation}
Using the value (\ref{int}) of the $M_0^J$ intercept, we find that the 
double baryon production cross section (\ref{cross2}) 
has $\sim s^{-1/2}$ energy dependence and (within the central rapidity 
region, where our considerations apply) does not depend on rapidity.  
The cross section (\ref{cross}) of single baryon stopping also decreases with 
energy, but much more slowly. Writing down the Pomeron intercept as
\begin{equation}
\alpha_P(0) = 1 + \Delta, \label{pomint}
\end{equation}
one gets the energy dependence of the cross section (\ref{cross}) 
in the form $\sim s^{-1/4 + \Delta/2}$. 
At very high energies the process of single baryon stopping will therefore be 
more important. Note however that the ISR data show a large value of 
the correlation between the probabilities of the stopping of the beam 
baryons, 
indicating that the process of double baryon stopping may still dominate 
even at rather high energies. 
The rapidity dependence of (\ref{cross}) does not show a ``central plateau", 
indicating instead a ``central valley" structure. This is in accord with 
the ISR data; moreover the particular rapidity dependence of (\ref{cross}) 
reproduces the data \cite{Cam} reasonably well.
\vskip0.3cm

The functional dependence of the single baryon stopping cross section 
similar to (\ref{cross}) 
has been advocated before \cite{Kop} in a different approach; the authors 
considered a specific mechanism of perturbative destruction of the 
fast diquark, 
accompanied by the baryon number flow over a large rapidity gap. 
In the framework of their approach, the authors of ref.\cite{Kop} have 
also performed a  
calculation of the cross section, based on the combination of perturbative 
technique and constituent quark model. 
\vskip0.3cm

Unfortunately, since we believe 
that the dynamics of the stopping process is genuinely non-perturbative, 
so far we have not been able to find 
a reliable way of computing the constant $G_p^M(0)$ entering the 
expression (\ref{cross}) and can only extract it from the existing data. 
However once it is done, we can perform parameter-free extrapolation 
to higher energies.  
\vskip0.3cm

As we have already stressed above, the formulae (\ref{cross2},\ref{cross})  
refer to the {\it net} baryon number, i.e. they refer to the difference 
between the number of produced baryons and antibaryons. 
The process of baryon-antibaryon pair production will therefore represent an 
important background to baryon stopping; the data \cite{Cam} 
show that at ISR 
energies the probability of baryon stopping is about three times 
smaller than the probability of baryon pair production. At high energies, the dominant 
contribution to the $\bar{B}B$ pair production will be given by the 
interaction of two Pomerons, with the cross section
\begin{equation}
E_B {d^3 \sigma^{(\bar{B}B)} \over d^3 p_B} = 
8 \pi [G_p^P(0)]^2\ f_{\bar{B}B}^{PP}(m_t^2)\ 
\left({\sqrt{s}\ m_t \over s_0}\right)^{2\alpha_P(0) - 2}, \label{pomfuse}
\end{equation} 
representing an energy-independent fraction of the total cross section:
\begin{equation}
\sigma^{\bar{B}B} \sim \sigma^{tot} \sim \left({s \over s_0}\right)^
{\alpha_P(0) - 1}.
\end{equation}
Since the topological structure of the Pomeron is that of a cylinder 
\cite{Ven}
\begin{equation}
P = Tr\left[ \left(P exp\left(ig\oint A_{\mu}dx^{\mu}\right)
\right) \right], 
\label{P}
\end{equation}
at high energies the associated multiplicities in the processes 
of single ($n^{(1)}$) 
and double ($n^{(2)}$)  
baryon stopping will be higher than in the average inelastic event, described 
by the cut of the Pomeron exchange diagram (see Fig.6):
\begin{equation}
n^{(1)} \simeq {5 \over 4}\ n^{inel};\hskip1cm 
n^{(2)} \simeq {3 \over 2}\ n^{inel}, \label{mult}
\end{equation}  
as we already discussed above at the qualitative level. Also, since 
the baryon stopping and baryon pair production arise in our scheme 
from different kinds of $t$-channel exchange ($M_0^J$ and the Pomeron, 
respectively), we expect a small correlation between the events with the pair 
production and stopping. Experimentally, it was found to be $0.16\pm0.22$ 
\cite{Cam}.
\vskip0.3cm

Before we turn to the discussion of nucleus-nucleus collisions, it is 
useful to recall the geometrical picture of high energy scattering in 
the impact parameter plane. In the impact parameter representation, the 
growth of the total cross section at high energies as described by 
one-Pomeron exchange can be attributed 
to the increase of the effective radius of the interaction, according to
\begin{equation}
R_{int} \simeq \sqrt{2 R_p^2 + \alpha_P'\ ln(s/s_0)},
\end{equation}    
where $R_p$ is a constant and 
$\alpha_P'$ is the slope of the Pomeron trajectory. 
(The Froissart bound allows even faster growth 
of the interaction radius with energy: $R_{int}\sim ln(s/s_0)$.) 
The central region of the disk becomes completely black at high energy if 
$\alpha_P(0)>1$; 
a further growth of the interaction strength in the centre of the 
disk is prevented by the unitarity constraint imposed on the partial 
amplitudes. 
\vskip0.3cm

The colliding nuclei, the transverse plane nucleon distributions in which are 
characterized by the packing parameter 
(\ref{r}) will therefore see each other as uniform black disks. 
This means that in a {\it central} nucleus-nucleus collision the  
cross section of the inelastic nucleon-nucleon collision will not 
further increase 
with energy when $R_{int}>> r$ since the 
soft peripheral interactions building up the Pomeron will be 
effectively screened out. On the other hand, the processes of baryon 
stopping are central in the impact parameter plane, and therefore  
may not be screened in the case of nuclear collisions. 
A very slow decrease of the cross section (\ref{cross}) with energy 
implies then that even at LHC energies the nuclear stopping may still 
be present, as we shall now discuss.
\vskip0.3cm

The ISR data \cite{Alp73,Ros75} show that at $\sqrt{s} = 53\ GeV$ 
the cross sections of proton and antiproton production 
at $y^*=0$ and $p_t=0.6\ GeV/c$ are 
\begin{equation}
{d^3\sigma^p \over d^3 p}(y^*=0) = 0.700\pm 0.162\ mb\ GeV^{-2};\ \ 
{d^3\sigma^{\bar{p}} \over d^3 p}(y^*=0) = 0.430\pm 0.033\ mb\ GeV^{-2}. 
\label{exp}
\end{equation}
Since the mechanism of baryon-antibaryon pair production (see (\ref{pomfuse})) 
obviously leads to an equal number of protons and antiprotons, the data 
(\ref{exp}) 
 imply that the following fraction of the protons in the central 
rapidity region is produced 
by stopping:
\begin{equation}
f_{st}(\sqrt{s}=53\ GeV) = {\sigma^p - \sigma^{\bar{p}} \over 
\sigma^p + \sigma^{\bar{p}}} \simeq 25\%. \label{stop}
\end{equation}
We can now estimate the ratio $R$ of multiplicities associated with proton 
($n^{p}$) and 
antiproton ($n^{\bar{p}}$) production, using the predictions (\ref{mult}). 
Since 
the multiplicity associated with protons and antiprotons should be the same 
in the absence of stopping, we get
\begin{equation}
R = {n^{p} \over n^{\bar{p}}} = (1-f_{st}) + f_{st}\ 
\left({n^{(1)} \over n^{\bar{p}}}\right), 
\label{rat}
\end{equation}
where we have omitted the contribution of the double stopping process, 
which is asymptotically suppressed at high energies according to 
(\ref{cross2}), but may still be important at ISR energies \cite{Cam}. 
Assuming that the multiplicity associated with antiprotons does not differ 
substantially from the average multiplicity of an inelastic event, 
$n^{\bar{p}} \simeq n^{inel}$, in accord 
with (\ref{pomfuse}) and with experimental data \cite{Bel77}, we obtain 
from (\ref{rat}) and (\ref{mult}) an estimate
\begin{equation}
R \simeq 1.05. \label{est}
\end{equation}
Even though the value (\ref{est}) agrees within experimental errors 
with the measured value of $R \simeq 1.1$ \cite{Bel77}, 
we may conclude that the contribution of double baryon stopping with higher 
associated multiplicity (\ref{mult}) is possible. A detailed experimental 
study 
of double baryon production in $pp$ collisions would therefore be useful 
to clarify the situation.
\vskip0.3cm

Extrapolation to the $pp$ collisions at the energies of 
$\sqrt{s}\simeq 6\ TeV$ (corresponding to the c.m.s. energy per nucleon-nucleon 
collision in Pb-Pb interactions at LHC) 
according to formulae (\ref{cross}, \ref{pomfuse}) 
with $\Delta = 0.08$ \cite{DL} yields then the following stopping fraction:
\begin{equation}
f_{st}(\sqrt{s} = 6\ TeV) \sim 5\%. \label{pp}
\end{equation}
The nucleus-nucleus collisions 
will be accompanied by much larger stopping, as we discussed above, 
and we expect that the estimate 
(\ref{pp}) in this case can only be considered as a lower bound on the baryon 
asymmetry. Therefore we may expect substantial excess of 
baryons over antibaryons 
in the central rapidity region of nucleus-nucleus collisions at LHC. 
\vskip0.3cm

It is interesting that already at SPS energy, the otherwise successful 
phenomenological approaches  
based on the topological expansion [13-16], 
%\cite{CT, Kai, Coh, CREV} 
but not taking 
into account the 
presence of the string junction explicitly, seem to underestimate \cite{Cap} 
the pronounced baryon stopping observed experimentally in nucleus-nucleus 
collisions \cite{stop}.  
\vskip0.3cm

It would be useful to analyse the dynamics of baryon stopping 
in an approach where the topological structure of the baryon is 
explicit: the Skyrme model. The formation of baryon-antibaryon pairs 
in this approach is treated as the formation of topological defects 
in the quark condensate \cite{Sky}  
(the net baryon number of the produced pairs is 
of course equal to zero). The picture proposed above would 
invoke into consideration also the high-energy 
scattering of such 
topological defects. Since the Lorentz boost of such configurations is not 
trivial, one may expect the occurence of the final states in which 
one or two Skyrmions are stopped in the central region, and the part of their 
pion field is ``shaken off" to the fragmentation region. 
Such events would contribute to the baryon asymmetry in the central 
region. We leave the consideration of 
the baryon stopping dynamics in topological models for further studies.  
\vskip0.3cm

Amazingly, the so-called ``Centauro"  and 
``Chiron" 
events reported in cosmic-ray emulsion experiments \cite{cos} 
and interpreted in favour 
of the existence of dense quark matter by Bjorken and McLerran \cite{BM},  
are characterized by non-vanishing baryon number density. 
As was formulated in ref. \cite{Bj}, ``the fluid in the central region has 
no net baryon number, so that there would need to be 
a spontaneous generation of net baryon density to make these objects".   
Indeed, in the scenario proposed by Bjorken \cite{Bj}, at ultra-relativistic 
energies the valence quarks of the colliding nuclei pass through each other, 
leaving behind a ``little Universe" of zero net baryon density, 
which at the moment 
of its production contains a gluon sea. In our 
picture, this sea from the very beginning is made stormy by the presence 
of non-perturbative twists - specific configurations of the gluon field, 
which trace non-zero {\it net} baryon number.  The ``little Universe", 
 just like our big one, may therefore generate substantial baryon 
asymmetry.
\vskip0.3cm

I am very indebted to G. Veneziano for many instructive suggestions and 
encouragement. 
It is a pleasure to acknowledge stimulating and 
enlightening discussions with 
J.-P. Blaizot, A. Capella, J. Ellis, K.J. Eskola, M. Ga\'zdzicki, K. Kajantie, 
L. McLerran, J.-Y. Ollitrault 
and H. Satz, whom I also thank for his interest in this study. 
This work was supported by the German Research Ministry (BMBW) under 
contract 06 BI 721.


\newpage

\begin{thebibliography}{999}

\bibitem{Ven}
{ G.C. Rossi and G. Veneziano, Nucl. Phys. {\bf B123} (1977) 507; Phys. Rep. 
{\bf 63} (1980) 153.} 
\bibitem{Ven1} 
{G. Veneziano, Nucl.Phys. {\bf B74} (1974) 365; 
Phys.Lett. {\bf B52} (1974) 220.}
\bibitem{Bel77}
{G. Belletini et al., Nuovo Cimento {\bf 42A} (1977) 85.}
\bibitem{Alp}
{B. Alper et al., Nucl. Phys. {\bf B100} (1975) 237.}
\bibitem{Ben}
{G. Bendiscioli and D. Kharzeev, Riv. Nuovo Cimento {\bf 17} (1994) No.6.}
\bibitem{Mue}
{A.H. Mueller, Phys. Rev. {\bf D2} (1970) 2963.}
\bibitem{Kan}
{O.V. Kancheli, JETP Lett. {\bf 11} (1970) 397.}
\bibitem{Cam}
{L. Camilleri, Phys. Rep. {\bf 144} (1987) 51.}
\bibitem{Kop}
{B.Z. Kopeliovich and B.G. Zakharov, Z. Phys. {\bf C43} (1989) 241.}
\bibitem{Alp73}
{B. Alper et al., Phys. Lett. {\bf B47} (1973) 275.}
\bibitem{Ros75}
{A.M. Rossi et al., Nucl. Phys. {\bf B84} (1975) 269.}
\bibitem{DL}
{A. Donnachie and P.V. Landshoff, Phys. Lett. {\bf B296} (1992) 227.}
\bibitem{CT}
{A. Capella and J. Tr\^an Th\^anh Van, Phys. Lett. {\bf B114} (1982) 450.}
\bibitem{Kai}
{A.B. Kaidalov, Phys. Lett. {\bf B116} (1982) 459;\\
A.B. Kaidalov and K.A. Ter-Martirosyan, Sov.J.Nucl.Phys. {\bf 39} (1984) 1545.}
\bibitem{Coh}
{G. Cohen-Tannoudji, A.E. Hassouni, J. Kalinowski and R. Peschanski, 
Phys. Rev. {\bf D19} (1979) 3397.}
\bibitem{CREV}
{A. Capella, U. Sukhatme, C.-I. Tan and J. Tr\^an Th\^anh Van, Phys. Rep. 
{\bf 236} (1994) 225.}
\bibitem{Cap}
{A. Capella, private communication.}
\bibitem{stop}
{See, for example,\\
The NA35 Collaboration, M. Ga\'zdzicki et al., Nucl. Phys. {\bf A590} 197c;\\
The NA49 Collaboration, S. Margetis et al.,  Nucl. Phys. {\bf A590} 355c.}
\bibitem{Sky}
{J. Ellis and H. Kowalski, Phys. Lett. {\bf B214} (1988) 161; Nucl. Phys. 
{\bf B327} (1989) 32;\\
J. Ellis, U. Heinz and H. Kowalski, Phys. Lett. {\bf B214} (1988) 161;\\
J.I. Kapusta and A.M. Srivastava,  Phys.Rev.{\bf D52} (1995) 2977;\\
see also T.A. DeGrand,  Phys.Rev.{\bf D30} (1984) 2001.}
\bibitem{cos}
{Brazil-Japan Emulsion Chamber Collaboration, {\it unpublished};\\
C.M.G. Lattes, Y. Fujimoto and S. Hasegawa, Phys.Rep. {\bf 65} (1980) 151.}
\bibitem{BM}
{J. Bjorken and L. McLerran, Phys.Rev.{\bf D20} (1979) 2353.}
\bibitem{Bj}
{J.D. Bjorken, Phys.Rev. {\bf D27} (1983) 140.}
\end{thebibliography}
\end{document}